\documentclass[final]{aipproc}
\usepackage{graphicx}
\usepackage{epsfig}
\usepackage{mathptm}
\layoutstyle{6x9}

\renewcommand{\theequation}{\arabic{equation}}

\newcommand{\pvalt}{\raise0.15ex\hbox{-}\mkern-11.5mu\int}
\newcommand{\be}{\begin{equation}}
\newcommand{\ee}{\end{equation}}
\newcommand{\bea}{\begin{eqnarray}}
\newcommand{\eea}{\end{eqnarray}}
\newcommand{\ben}{\begin{enumerate}}
\newcommand{\een}{\end{enumerate}}
\newcommand{\bit}{\begin{itemize}}
\newcommand{\eit}{\end{itemize}}
\newcommand{\0}{0^{++}}
\newcommand{\la}[1]{\label{#1}}

\newcommand{\eq}[1]{eq.~(\ref{#1})}

\newcommand{\half}{\frac{1}{2}}

\newcommand\qq{\bar q q}
\newcommand\qqqq{\bar q^{2}q^{2}}

\newcommand{\de}{\delta}

\renewcommand{\O}{{\cal O}}
\newcommand{\fm}{{\rm fm}}
\newcommand{\GeV}{{\rm GeV}}
\newcommand{\MeV}{{\rm MeV}}

\newcommand{\beq}{\begin{equation}}
\newcommand{\eeq}{\end{equation}}
\newcommand{\ba}{\begin{array}}
\newcommand{\ea}{\end{array}}
\newcommand{\dsp}{\displaystyle}

\renewcommand{\>}{\rangle} 
\newcommand{\exotic}{{\rm E}}
\newcommand{\nonexotic}{{\rm N}}
\newcommand{\dz}{\de E_\nonexotic}
\newcommand{\dt}{\de E_\exotic}
\newcommand{\etal}{{et~al.}}

\begin{document}


\title{Scalar Mesons as $\qqqq$?\\
Insight from the Lattice\footnote{Talk presented by R.~L.~Jaffe at the Scalar
Meson Workshop, May 2003, SUNYIT, Utica, NY}}

\author{M.~Alford}{
   address={Physics Department\\
   Washington University CB 1105\\
   St.~Louis, Missouri 63130},
}
\author{R.~L.~Jaffe}{
address={Center for Theoretical Physics,\\
Laboratory for Nuclear Science and Department of Physics\\
Massachusetts Institute of Technology, \\Cambridge, Massachusetts 02139},
}

\begin{abstract}
  I describe some insight obtained from a lattice calculation on the
possibility that the light scalar mesons are $\qqqq$ states rather
than $\qq$.  First I review some general features of $\qqqq$ states in
QCD inspired quark models.  Then I describe  a lattice QCD calculation of
pseudoscalar meson scattering amplitudes, ignoring quark loops and
quark annihilation, which finds indications that for sufficiently heavy
quarks there is a stable four-quark bound state with $J^{PC}=0^{++}$
and non-exotic flavor quantum numbers.
\end{abstract}

\maketitle
\newpage

\section{Introduction}
\label{section0}

We would like to call your attention to some work that we
performed a few years ago, that bears on the interpretation of the
scalar mesons. A complete description of our work can be found in
Ref.~\cite{Alford:2000mm}.

The light scalar mesons have defied classification for decades
\cite{Caso:1998,Black:1999}.  Some are narrow and have been firmly
established since the 1960's.  Others are so broad that their very
existence is controversial.  Scalar mesons are predicted to be chiral
partners of the pseudoscalars like the pion, but their role in chiral
dynamics remains obscure.  Naive quark models interpret them as
orbitally excited $\qq$ states.  Others have suggested that they are
$\qqqq$\cite{Jaffe:1977} or ``molecular'' states,\cite{Weinstein:1990}
strongly coupled to $\pi\pi$ and $\bar K K$ thresholds.  Recently a
consensus has emerged (at least in some quarters) that the light
scalar mesons have important $\qqqq$ components at short distances and
important meson-meson components at long distances.\cite{Close:2002zu}

We propose a new way to shed some light on the nature of the
scalar mesons using lattice QCD. Previously scalar mesons have been
treated like other mesons: their masses have been extracted from the
large Euclidean time falloff of $\qq-\qq $ correlation functions with
the appropriate quantum numbers.  Here we look for a $\0\,\qqqq$
\emph{bound state}.  We construct $\qqqq$ sources, work in the
quenched approximation, and discard $\qq$ annihilation diagrams so
communication with $\qq$ and vacuum channels is forbidden.  Also, we
allow the quark masses to be large (hundreds of MeV), so the continuum
threshold for the decay $\qqqq\to (\qq) (\qq)$ is artificially
elevated.  We then study the large Euclidean time falloff of a
$\qqqq-\qqqq$ correlator, looking for a falloff slower than
$2m_{\qq}$, signalling a bound state.  Such an object would have been
missed by studies of $\qq$ correlators in the quenched approximation.
We use shortcomings of lattice QCD to our advantage.  By excluding
processes that mix $\qq$ and $\qqqq$, we can unambiguously assign a
quark content to a state.  The heavy quark mass suppresses
relativistic effects, which we believe complicate the interpretation
of light quark states.

Our initial results are encouraging: within the limits of our
computation we see signs of a bound state in the ``non-exotic''
$\qqqq$ channel, the one with quantum numbers that could also
characterize a $\qq$ state ($I=0$ for 2 flavors, the {\bf 1} and {\bf
8} for 3 flavors).  In contrast, the ``exotic'' flavor $\qqqq$ channel
($I=2$ for 2 flavors, the {\bf 27} for 3 flavors) shows no bound
state.  Instead it shows a negative scattering length, characteristic
of a repulsive interaction.  A definitive result will
require larger lattices and more computer time, but this is well
within the scope of existing facilities.
There have been a couple of previous studies of $\qqqq$ sources on
the lattice.\cite{GPS2,Fukugita} Because these earlier works looked
only at one (relatively small) lattice size they were unable to
examine the possibility of a bound state.  

A reader who wishes to skip the details can look immediately at
Fig.~\ref{fig:falloff} where we plot the dependence on lattice size of
the binding energy of the exotic and non-exotic $\qqqq$
channels.  The exotic channel shows a negative binding energy with the
$1/L^{3}$ dependence expected from analysis of the $(\qq)(\qq)$
continuum.\cite{Luscher} The coefficient of $1/L^{3}$ agrees roughly
with Refs.~\cite{GPS2,Fukugita} and with the predictions of chiral
perturbation theory.  The non-exotic channel shows positive binding
energy, but seems to depart from $1/L^{3}$, perhaps approaching a
constant as $L\to\infty$, which would indicate the existence of a
bound $\qqqq$ state.  Confirmation of this result will require
calculations on larger lattices.

\section{Overview of the light scalar mesons}
\label{section1}

In this section we give a very brief introduction to the phenomenology
of the lightest $\0$ mesons composed of light ($u$, $d$, and $s$)
quarks and existing lattice calculations.

The known $\0$ mesons divide into effects near and below 1 GeV,
which are unusual, and effects in the 1.3--1.5 GeV region which may be
more conventional.  Here we focus on the states below 1 GeV.
Altogether, the objects below $1~\GeV$ form an $SU(3)_{\rm f}$ nonet: two
isosinglets, an isotriplet and two strange isodoublets.  The
isotriplet and one isosinglet are narrow and well confirmed.  The
isodoublets and the other isosinglet are very broad and still
controversial.

The well established $0^{++}$ mesons are the isosinglet $f_{0}(980)$
and the isotriplet $a_{0}(980)$.  Both are relatively narrow:
$\Gamma[f_{0}]\sim$ 40 MeV, $\Gamma[a_{0}]\sim$ 50 MeV,\footnote{We
use the observed peak width into $\pi\pi$ and $\pi\eta$ respectively,
rather than some more model dependent method for extracting a width.}
despite the presence of open channels ($\pi\pi$ for the $f_{0}$ and
$\pi\eta$ for the $a_{0}$) for allowed s-wave decays.   Both couple
strongly to $\bar K K$ and lie so close to the $\bar K K$ threshold
at 987 MeV that their shapes are strongly distorted by threshold
effects.  Interpretation of the $f_{0}$ and $a_{0}$ requires a
coupled channel scattering analysis.  The relevant channels are
$\pi\pi$ and $\bar K K$ for the $f_{0}$ and $\pi\eta$ and $\bar K K$
for the $a_{0}$.  In both cases the results favor an intrinsically
broad state, strongly coupled to $\bar K K$ and weakly coupled to the
other channel.  The physical object appears narrow because the $\bar
K K$ channel is closed over a significant portion of the object's width.
No summary this brief does justice to the wealth of work and
opinion in this complex situation.

The other light scalar mesons are known as broad enhancements in very
low energy s-wave meson-meson scattering.  The enhancements are
universally accepted, but their interpretation is more controversial.
At the lowest energies only the $\pi\pi$ channel is open.  The
$\pi\pi$ s-wave can couple either to isospin zero or two.  The $I=2$
(e.g.  $\pi^{+}\pi^{+}$) channel shows a weak repulsion in rough
agreement with the predictions of chiral low energy
theorems.\cite{Weinberg} The $I=0$ channel shows a strong
attraction: the phase shift rises steadily from threshold to
approximately $\pi/2$ by $\sim 800$ MeV before effects associated with
the $f_{0}$ complicate the picture.  This low mass enhancement in the
$\pi\pi$ s-wave is the $\sigma$ meson of nuclear physics and chiral
dynamics.  Recent studies support the existence of an S-matrix pole
associated with this state at a mass around 600 MeV, which we will
refer to as the $\sigma(600)$.\cite{Black:1999,Tornqvist:1996ay} The $\pi
K$ s-wave is very similar to $\pi\pi$.  The exotic $I=3/2$ (e.g.
$\pi^{+} K^{+}$) channel shows weak repulsion.  The non-exotic $I=1/2$
channel shows relatively strong attraction.

The conventional quark model assigns the $\0$ mesons to the first
orbitally excited multiplet of $\qq$ states.  As in positronium, $\0$
quantum numbers are made by coupling $L=1$ to $S=1$ to give total
$J=0$.  The $\0$ states should be very similar to the $1^{+\pm}$ and
$2^{++}$ $\qq$ states that lie in the same family.  These are very
well known and form conventional meson nonets (in $SU(3)_{\rm f}$).
Since they have a unit of excitation (orbital angular momentum), they
are expected to be quite a bit heavier than the pseudoscalar and
vector mesons.  Most models put the $\qq~\0$ mesons along with
their $2^{++}$ and $1^{++}$ brethren around 1.2--1.5 GeV.

\begin{figure}
\includegraphics[width=6.5in]{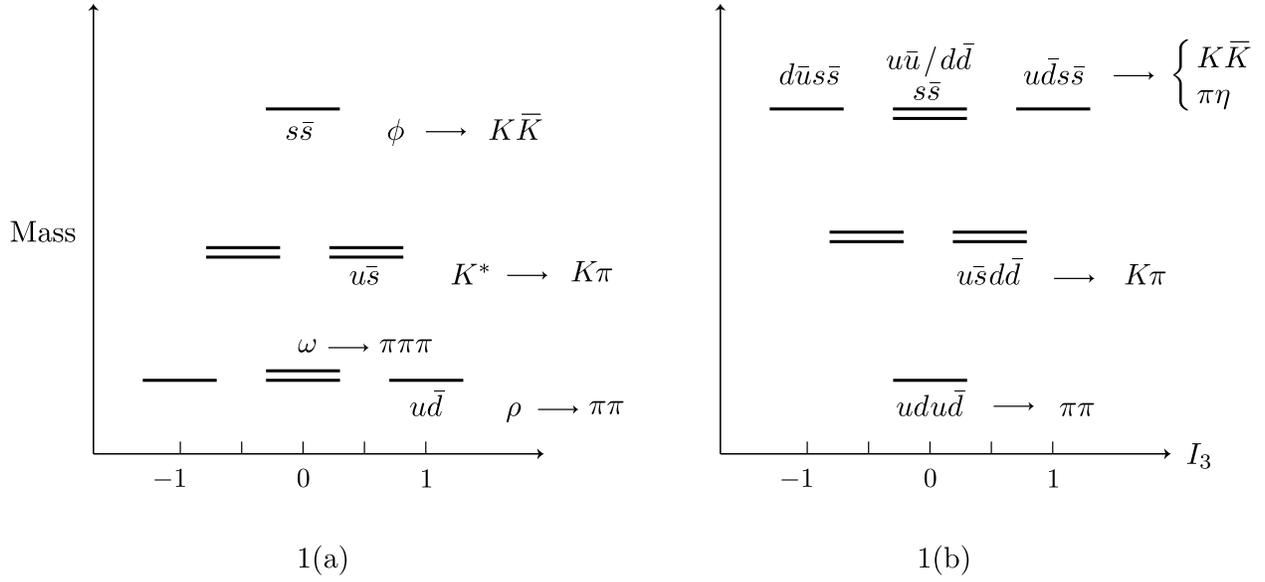}
\vspace{3mm}
\caption{The mass pattern, quark
content and natural decay couplings of (a) a $\qq$ nonet
and (b) a $\qqqq$ nonet.}
\label{quark}
\end{figure}

An idealized $\qq$ meson nonet has a characteristic pattern of
masses and decay couplings.  The vector mesons are best known, but the
pattern is equally apparent in the $2^{++}$ or $1^{++}$ nonets.  The
isotriplet and the isosinglet composed of non-strange quarks are
lightest and are roughly degenerate (e.g.~the $\rho$ and the $\omega$).
The strange isodoublets are heavier because they contain a single
strange quark (e.g.~the $K^{\ast}$).  The final isosinglet is heaviest
because it contains an $\bar s s$ pair (e.g.~the $\phi$).  Decay
patterns show selection rules which follow from this quark content.
In particular, the lone isosinglet does not couple to non-strange
mesons ($\phi\not\!\to 3\pi$).
The mass pattern, quark
content and natural decay couplings of a $\qq$ nonet are summarized in
Fig.~\ref{quark}a.
These patterns seem to bear  little
resemblance to the masses and couplings of the light
$\0$ mesons, a fact which led earlier workers to explore other
interpretations.

Four quarks ($\qqqq$) can couple to $\0$ without a unit of orbital
excitation.  Furthermore, the color and spin dependent interactions,
which arise from one gluon exchange, favor states in which quarks and
antiquarks are separately antisymmetrized in flavor.  For quarks in
3-flavor QCD the antisymmetric state is the
flavor $\bar {\bf 3}$.  Thus
the energetically favored configuration for $\qqqq$ in flavor is
$(\bar q\bar q)^{{\bf 3}}(qq)^{\bar {\bf 3}}$,
a flavor nonet.
The lightest multiplet has spin 0.  Explicit studies in the MIT Bag
Model indicated that the color-spin interaction could drive the
$\qqqq$ $\0$ nonet down to very low energies: 600 to 1000 MeV
depending on the strangeness content.\cite{Jaffe:1977}

The most striking feature of a $\qqqq$ nonet in comparison with a
$\qq$ nonet is an \emph{inverted mass spectrum}
(see Fig.~\ref{quark}b).  The crucial
ingredient is the presence of a hidden $\bar s s$ pair in several
states.  The flavor content of $(qq)^{\bar {\bf 3}}$ is $\{[ud],
[us], [ds]\}$, where the brackets denote antisymmetry.  When combined
with $(\bar q\bar q)^{{\bf 3}}$, four of the resulting states
contain a hidden $\bar s s$ pair: the isotriplet and one of the
isosinglets have quark content $\{u\bar d s \bar s,
\frac{1}{\sqrt{2}}(u\bar u-d\bar d)s\bar s, d\bar u s \bar s\}$ and
$\frac{1}{\sqrt{2}}(u\bar u+d\bar d)s\bar s$, and therefore lie at the
top of the multiplet.  The other isosinglet, $u\bar d d \bar u$ is the
only state without strange quarks and therefore lies alone at the
bottom of the multiplet.  The strange isodoublets ($u\bar s d\bar d$,
etc.)  should lie in between.  In summary, one expects a degenerate
isosinglet and isotriplet at the top of the multiplet and strongly
coupled to $\bar K K$, an isosinglet at the bottom, strongly coupled to
$\pi\pi$, and a strange isodoublet coupled to $K\pi$ in between
(Fig.~\ref{quark}b).  The resemblance to the observed structure of the
light $\0$ states is considerable.

These qualitative considerations motivate a careful look at the
classification of the scalar mesons.  Models of QCD are not
sophisticated enough to settle the question.  For example, the $\qqqq$
picture does not distinguish between one extreme where the four quarks
sit in the lowest orbital of some mean field,\cite{Jaffe:1977} and the
other, where the four quarks are correlated into two $\qq$ mesons
which attract one another in the flavor $(\bar q\bar q)^{{\bf 3}}
(qq)^{\bar {\bf 3}}$ channel.\cite{Jaffe:1979bu,Weinstein:1990} For
years, phenomenologists have attempted to analyse meson-meson
scattering data in ways which might distinguish between $\qq$ and
$\qqqq$ assignments.  A recent quantitative study favors the $\qqqq$
assignment.\cite{Black:1999} However the $\qq$ assignment has strong
advocates.  We hope that a suitably constrained lattice calculation
can aid in the eventual classification of these states.

Only a very few lattice calculations
bear on the classification of the $\0$ mesons.
There have been lattice studies of both the spectrum of $\0$
states and the mixing of $\qq$ states with glueballs.

Unquenched spectroscopic calculations are just beginning to become
available\cite{SCALAR,UKQCD2}.  In principle, these are of interest
because they would couple to a $\qqqq$ configuration if it is
energetically favorable. Both these studies find that the mass
of the $\0$ state is lower than that reported in
quenched calculations.  We return to this work briefly
in our conclusions.  Further insight from unquenched calculations will
have to await more definitive studies.

Quenched calculations of the spectrum of $\qq$ states find the $\0$
states roughly degenerate with the other positive parity states.  Their
masses cluster around 1.3-1.5 times the $\rho$ mass and are relatively
constant as the ratio $m_{\pi}/m_{\rho}$ is reduced toward the chiral
limit.  In short they behave like other $\qq$ mesons.  For more details
the reader should consult Ref.~\cite{Alford:2000mm}.

In the past, lattice studies of four-quark states have been undertaken
only in order to extract pseudoscalar-pseudoscalar ($P$-$P$)
scattering lengths for comparison with the predictions of chiral
dynamics.  It is known \cite{Luscher} that the energy shift $\de E$ of
a two-particle state with quantum numbers $\alpha$ in a cubic box of
size $L$ is related to the threshold scattering amplitude,
\beq
\label{luscher}
\de E_{\alpha} = E_{\alpha} - 2 m_P =
\frac{T_{\alpha}}{L^3}\left(  1 + 2.8373\, \frac{m_P T_{\alpha}}{4 \pi L}
+ 6.3752 \left(\frac{m_P T_{\alpha}}{4 \pi L}\right)^{\!2}
  + \cdots \right),
\eeq
where $m_P$ is the mass of the scattering particles, and $T_{\alpha}$ is
the scattering amplitude at threshold
in the channel labelled by $\alpha$, which can be related to
the scattering length,
\be
T_{\alpha}=-\frac{4\pi a_{\alpha}}{m_P}.
\la{scatt-length}
\ee
For a more detailed discussion, see Ref.~\cite{GPS1}. In our case the
channels of interest are exotic ($I=2$, for two flavors) and non-exotic
($I=0$, for two flavors).  If the interaction is attractive enough to
produce a bound state, then instead of \eq{luscher} one would find that
$\de E$ goes to a negative constant as $L\to \infty$.

In order to distinguish between a bound state and the continuum
behavior described by \eq{luscher}, it is necessary to perform
calculations for several different lattice sizes.  Calculations with
$\qqqq$ sources have been performed by Gupta~\etal \cite{GPS2}, who
studied one lattice volume at one lattice spacing, and Fukugita~\etal
\cite{Fukugita}, who, for the heavy quark masses we are interested in,
also studied only one lattice volume at one lattice spacing.  Their
results were therefore not sufficient to check the lattice-size
dependence of the energy of the two-pseudoscalar state,
and investigate the possibility of a bound state.
Our method follows theirs, but we have studied a range of lattice sizes.
Their results are plotted along with ours in
Fig.~\ref{fig:falloff}.  Where our calculations overlap, they agree.

\section{A ${\bf \qqqq}$ Exercise on the Lattice}
\label{section2}

\begin{figure}
\includegraphics[width=5.5in]{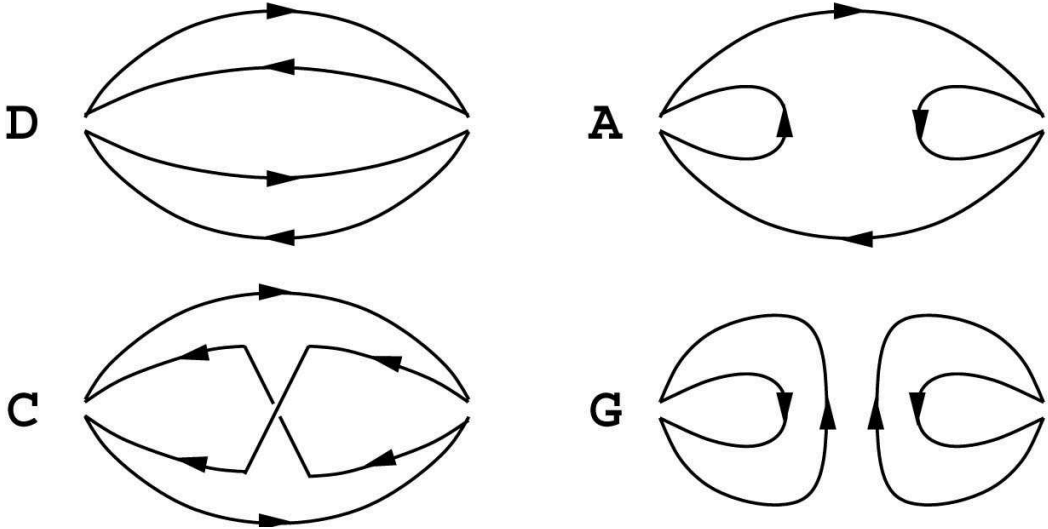}
\vspace{3ex}
\caption{ The four types of quark line contraction that contribute to
the pseudoscalar-pseudoscalar ($P$-$P$) correlation function.
}
\label{fig:dcag}
\end{figure}

For our purposes the salient categorization of
$\qqqq$ correlators is into ``exotic''
channels (flavor states that are only possible for a $\qqqq$ state,
$I=2$ for two flavors, the {\bf 27} for three flavors) and
non-exotic channels (flavor states that could be $\qqqq$ or $\qq$,
$I=0$ for two flavors, the {\bf 8} and {\bf 1} for three flavors).
In the absence of quark annihilation diagrams, the {\bf 8} and {\bf 1}
are identical. When annihilation is included, the  {\bf 1}, like the
$I=0$ for two flavors, can mix with pure glue.
As shown in Fig.~\ref{fig:dcag}, the $\qqqq~\0$ correlation functions can be
expressed in terms of a basis determined by the
four ways of contracting the quark
propagators\cite{GPS1}: direct (D), crossed (C), single annihilation (A),
complete annihilation into glue (G).
Since we are interested in $\qqqq$ states, we only study the D and C
contributions.  We will assume that all quarks are degenerate, so
there is only one quark mass, and as far as color and spinor indices
are concerned all quark propagators are the same. In our lattice
calculation we will therefore build our $\qqqq$ correlators from color
and spinor traces of contractions of four identical quark
propagators, putting in the flavor properties by hand when we choose
the relative weights of the different contractions.

In the case of two flavors, there are two possible channels
for a spatially symmetric source: $I=2$ (exotic) and $I=0$ (non-exotic).
Evaluation of the flavor dependence of the quark line contractions shows that
the $I=2$ channel is $D-C$, and $I=0$ is $D+\half C$ \cite{GPS1}.

For three flavors, the possible channels are the symmetric parts of
${\bf 3}\times{\bf 3}\times\bar{\bf 3}\times\bar{\bf 3}$, namely ${\bf
1}+{\bf 8}$ (non-exotic) and ${\bf 27}$ (exotic).  As in the
two-flavor case, the exotic channel is $D-C$.  At sufficiently large
Euclidean time separation, each contraction will behave as a sum of
exponentials, corresponding to the states it overlaps with.
Generically, all linear combinations will be dominated by the same
state: the lightest.  Only with correctly chosen relative weightings
will the leading exponential cancel out, yielding a faster-dropping
exponential corresponding to a more massive state.  The exotic ($D-C$)
channel is the one where such a cancellation occurs, yielding a
repulsive interaction between the pseudoscalars.  For {\em any} other
linear combination of $D$ and $C$ the correlator is therefore
dominated by the lightest, attractive state.  Without loss of
generality, we can therefore study the following linear combinations:
\beq \ba{r@{\qquad}rcl@{\qquad\mbox{2 flavor:~}}l@{\qquad\mbox{3
flavor:~}}l } \mbox{Exotic:} & J_\exotic &=& D - C & I=2 & {\bf 27} \\
\mbox{Non-exotic:} & J_\nonexotic &=& D + \half C  & I=0 & {\bf 1}, {\bf 8}
\ea
\eeq

We conclude that if, as our results suggest, there is a bound $\qqqq$ state
in the non-exotic channel, then this means that with two flavors,
the $I=0$ channel is bound, and with three flavors both the
{\bf 1} and {\bf 8} are bound.
Once quark loops and annihilation diagrams are included, the
{\bf 1} and {\bf 8} will split apart. Unquenched lattice calculations
will be needed to see if they remain bound.

In our lattice calculations, we work in the quenched (valence)
approximation, and use Symanzik-improved glue and quark actions.  This
means that irrelevant terms ($\O(a), \O(a^2)$, where $a$ is the
lattice spacing) have been added to the lattice action to compensate
for discretization errors.  Improved actions are crucial to our
ability to explore a range of physical volumes using limited computer
resources.  Because most of the finite-lattice-spacing errors have
been removed, we can use coarse lattices, which have fewer sites and
hence require much less computational effort: note that the number of
floating-point-operations required even for a quenched lattice QCD
calculation rises faster than $a^{-4}$.

Improved actions have been studied extensively
\cite{land234,WW,LW,Alf1,Improving}, and it has been found that even
on fairly coarse lattices ($a$ up to $0.4~\fm$) good results can be
obtained for hadron masses by estimating the coefficients of the
improvement terms using tadpole-improved perturbation theory.  For the
energy differences that we measure, we find that the improved action
works very well.  There are no signs of lattice-spacing dependence at
$a$ up to $0.4~\fm$, so as well as greatly reducing the computer
resources required, it enables us to dispense with the extrapolation
in $a$ that is usually needed to obtain continuum results.

We work at a quark mass close to the physical strange quark: the
pseudoscalar to vector meson mass ratio $m_P/m_V$ is 0.76.  We
emphasize that this is not entirely unwelcome, since it makes our
results easier to interpret.

To obtain the binding energy $\dz$ in the $I=0$ channel,
and the binding energy $\dt$ in the $I=2 $ channel, we
construct ratios of correlators and fit them to an exponential
\beq
\label{fitform}
\ba{c} \dsp R_\nonexotic(t) = \frac{J_\nonexotic(t)}{\<P(t)\>^2} \sim
A \exp(-\dz t), \\[2ex] \dsp R_\exotic(t) =
\frac{J_\exotic(t)}{\<P(t)\>^2} \sim B \exp(-\dt t).  \ea \eeq Here
$J_{\nonexotic}$ and $J_{\exotic}$ are the $D+\frac{1}{2}C$
(non-exotic) and $D-C$ (exotic) correlators respectively, and $P$ is
the pseudoscalar correlator.  $t$ is the Euclidean lattice time.  The
ratios of correlators are expected to take the single exponential form
only at large $t$, after contributions from excited states have died
away.  We followed the usual procedure of looking for a plateau and
found no difficulty in identifying the
plateau and extracting $\dz$ or $\dt$.

Since this is not a lattice workshop I will spare you further details.
However the reader can find a discussion of the sources we used and
our fitting methods in Ref.~\cite{Alford:2000mm}.

\section{Results and Discussion}
\label{section3}

\begin{figure}
\includegraphics[width=4.5in]{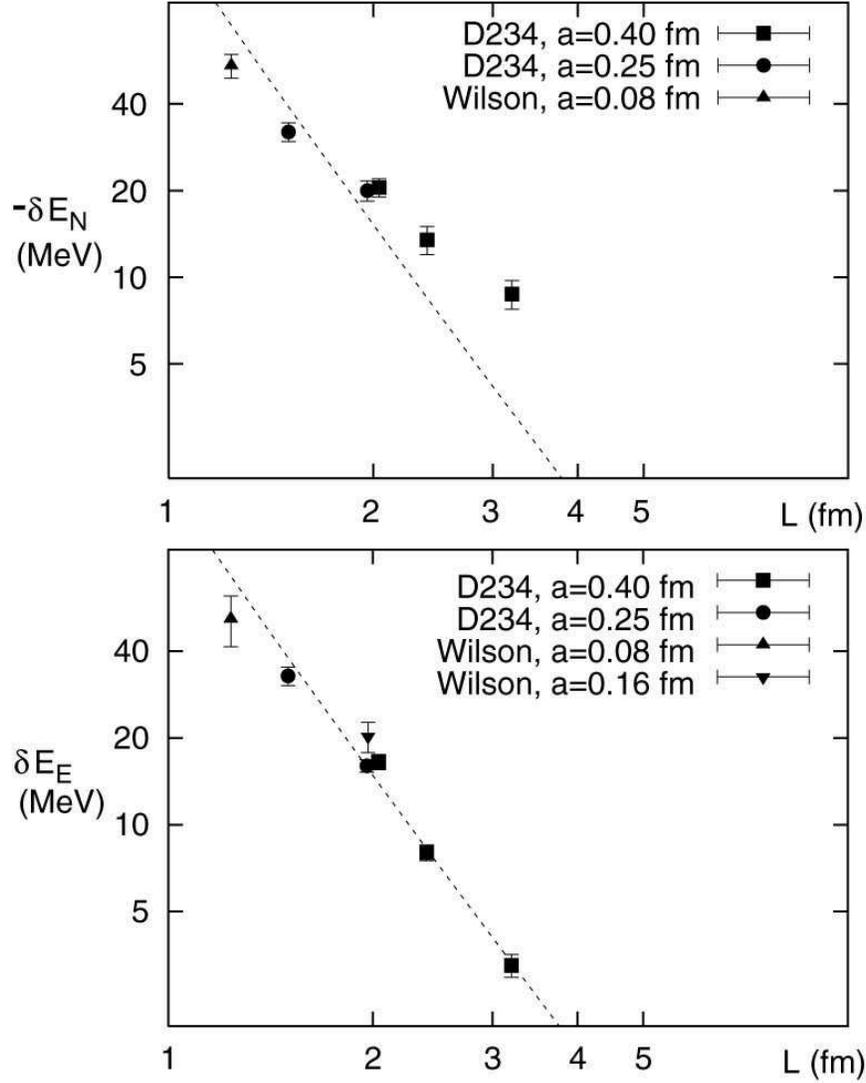}
\caption{ $P$-$P$ binding energy in non-exotic (N) and exotic
(E) channels.
The data at $a=0.08~\fm$ are from \protect\cite{GPS2};
the data at $a=0.16~\fm$ are from \protect\cite{Fukugita}.
The $a=0.25~\fm$ and $a=0.4~\fm$ points at $L=2~\fm$
have been displaced slightly to either side in order to distinguish them.
The lines are fits to \protect\eq{luscher}.
}
\label{fig:falloff}
\end{figure}

We measured $\delta E_{N}$ and $\delta E_{E}$ for several different
lattice spacings and sizes.  Our results are shown in
Fig.~\ref{fig:falloff} along with previous results from
Refs.~\cite{GPS2,Fukugita}.  The exotic and non-exotic channels appear
to scale differently as a function of $L$.  The exotic channel falls
like $1/L^3$, which is the expected form for a scattering state,
\eq{luscher}.  A fit is shown in the figure.  The non-exotic channel
appears to depart from $1/L^3$ falloff.  To be complete, however, we
have fitted the non-exotic data also to the form expected for a
scattering state.

Our results are consistent with those of Refs.~\cite{GPS2,Fukugita}, even
though we use much coarser lattices.  This supports our use of
Symanzik-improved glue and quark actions with tadpole-improved
coefficients.  As a further check on the validity of the improved
actions, we note that at $L=2~\fm$, where we performed a calculation at
two different lattice spacings for the same lattice volume, the
results for the two lattice spacings agree very well.  There is no
evidence of any discretization errors.

For the exotic $\qqqq$ system, the fit to \eq{luscher} is quite
good, and the fitted scattering amplitude is remarkably similar to the result
expected in the chiral limit, $4f_P^2 T=1$.\footnote{Since we did not
calculate $f_P$ at our quark masses, we have used the value $f_P =
148$ MeV, derived from Ref.~\cite{GPS2}, Table 1.}  We conclude that
there are no surprises in the exotic channel -- the interaction near
threshold appears repulsive and the strength is close to that
predicted by chiral perturbation theory.

The non-exotic $\qqqq$ system, however, does not fit the expected
scaling law at large $L$.  The fit to \eq{luscher} has a very large
$\chi^2$, and is so poor that the extracted amplitude $T$ is
meaningless.  Instead $\dz$ appears to be approaching a negative
constant at large $L$.  Instead of a scattering state, we appear to be
seeing a \emph{bound state} in the non-exotic channel.  Although our
data are suggestive, they are not conclusive.  It would be very
interesting to gather more data at $L\sim 4$ fm, as well as at a
range of quark masses, in order to verify the existence of this new
state in the quenched hadron spectrum.

Apparently we have evidence for a $\qqqq$ bound state just below
threshold in the non-exotic pseudoscalar-pseudoscalar $s$-wave.  In
2-flavor QCD the bound state would correspond to an isosinglet meson
coupling to $\pi\pi$.  In 3-flavor QCD the non-exotic channel
corresponds to an entire nonet including two non-strange isosinglets
and an isotriplet, and two strange isodoublets (see
Fig.~\ref{quark}b).  We work with a large quark mass so our results
are not directly applicable to $\pi\pi$ scattering, but they do
resemble physical $K\bar K$ scattering.\footnote{Although we work in
the $SU(3)_{\rm f}$ limit where all quark masses are equal.} The known
isosinglet $f_{0}(980)$ and isotriplet $a_{0}(980)$ mesons are obvious
candidates to identify with the non-exotic $\qqqq$ bound states we
seem to have found on the lattice.

We believe the quark mass dependence of the non-exotic $\qqqq$ state
is quite different from a standard $\qq$ lattice state.  In the
quenched approximation the masses of $\0\qq$ states have been found to
be roughly independent of $m_{P}$.  At large quark mass the $\qq\ \0$ mass
is below $2m_{P}$, but as $m_{P}$ is decreased the $\qq\ \0$ mass crosses
threshold, $2m_{P}$.  It appears to be smooth as it crosses the
threshold.  In contrast, we believe that the $\qqqq$ state
we may have identified is strongly correlated with the $PP$ threshold
when the quark mass is large, and departs from it in a characteristic
way as the quark mass is reduced.  (Indirect support for this comes
from Gupta~\etal's finding that their binding energy is independent of
the pseudoscalar mass.)  In particular, we believe that the bound
state will move off into the meson-meson continuum as $m_{P}$ is
reduced toward the physical pion mass.

\begin{figure}
\includegraphics[width=4.5in]{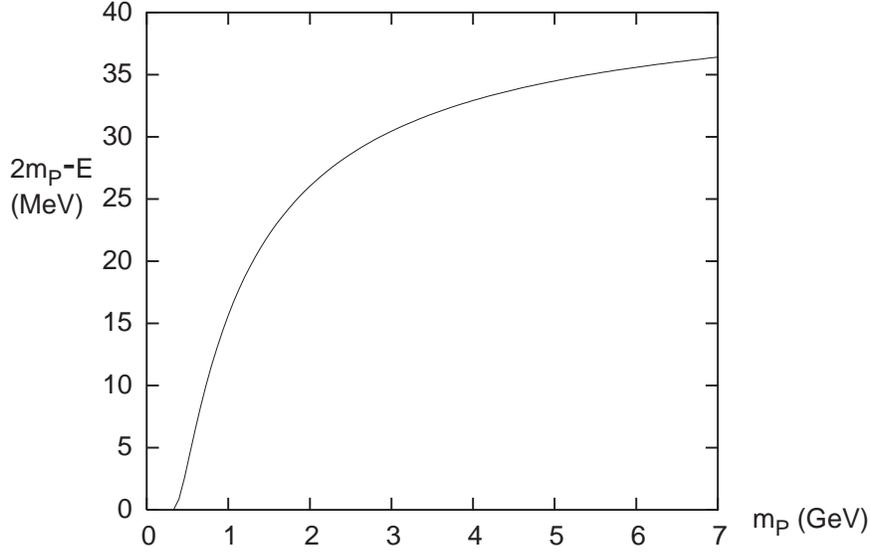}
\vspace{3mm}
\caption{Binding energy in MeV of the two-particle
state in our toy model \eq{toy}, as a function of particle mass in GeV.}
\label{toymodel}
\end{figure}

To explore the $m_{P}$ dependence of our results, we have made a toy
model based on a relativistic generalization of potential scattering.
We write a Klein-Gordon equation for the $s$-wave relative meson-meson
wavefunction, $\phi(r)$,
\be\label{toy}
    -\phi^{\prime\prime}(r) + (2m_{P}-U(r))^{2}\phi(r)= E^{2}\phi(r),
    \label{kg}
\ee
with the boundary condition that $\phi(0)=0$.
For $U(r)=0$ the spectrum is a continuum beginning at $E=2m_{P}$ as
required.  In the non-relativistic limit $m_{P}\ll |U|$,
\eq{kg} reduces to the
Schr\"odinger equation with an attractive potential $-U(r)$ (for
$U(r)>0$).  For sufficient depth and range, this potential will have a
bound state.  However,
as $m_{P}\to 0$, the potential term in \eq{kg} turns repulsive and
the bound state disappears.  Thus, if one keeps the depth and range of
$U$ fixed as one decreases $m_{P}$, the bound state moves out into the
continuum and disappears.  To be quantitative, we have taken a square
well, $U(r)=U_{0}$, for $r\le b$, and $U(r)=0$ for $r>b$.  We chose a
range $b=1/m_{\pi}\approx$ 1.4 fm, and adjusted $U_{0}$ such that
the bound state has binding energy of 10 MeV when $m_{P}\sim800~\MeV$.
The bound state does indeed move off into the continuum (first as a
virtual state) when $m_{P}$ goes below $330~\MeV$. The behavior of the
bound state in this toy model is shown in Fig.~\ref{toymodel}.  Note
this toy model is not meant to be definitive%
\footnote{
We could have chosen a different relativistic
generalization of the Schr\"odinger equation which would have
preserved the bound state
as $m_{P}\to 0$.  For example, we could have replaced
$(2m_{P}-U)^{2}$ by $2m_{P}^{2}-2m_PU_1-U_2^2$, and fine-tuned
$U_1$ and $U_2$ to provide binding at arbitrarily low $m_P$.
}
but it illustrates the expected behavior of a $P$--$P$ bound state:
tracking $2m_P$ with roughly constant binding energy as $m_P$ falls, then
unbinding at some critical  $m_P$.

On the basis of our lattice computation and the $m_{P}$ dependence
suggested by our toy model, we believe it is possible that \emph{all}
the phenomena associated with the light scalar mesons are linked to
$\qqqq$ states.  The narrow $\0$ isosinglet $f_{0}(980)$ and
isotriplet $a_{0}(980)$ mesons near $K\bar K$ threshold can be
directly identified with $\qqqq$ lattice bound states (top line of
Fig.~\ref{quark}b).  The broad $\kappa(900)$ and $\sigma(600)$ (middle
and bottom lines of Fig.~\ref{quark}b) couple to low mass ($\pi\pi$ or
$\pi K$) channels.  We speculate that they are to be identified as the
continuum relics of the same objects which appear as bound states of
heavy quarks.

Of course, a thorough examination of this question would require
implementing flavor $SU(3)$ violation by giving the strange quark a
larger mass.  This would mix and split the isoscalars, shift the other
multiplets (see Fig.~\ref{quark}b), and dramatically alter
thresholds. For example, the $I=1$ $\qqqq$ state couples both to
$K\bar K$ and $\pi\eta$ (through the $\bar s s$ component of the
$\eta$) in the quenched approximation. The fact that the physical
$K\bar K$ and $\pi\eta$ thresholds are significantly different would
certainly affect the manifestation of bound states such as those
we have been discussing in the $SU(3)$-flavor-symmetric limit.

In summary we have presented evidence for previously unknown
pseudoscalar meson bound states in lattice QCD. Our results need
confirmation.  Calculations on larger lattices are needed, and
variation with quark mass, lattice spacing, and discretization scheme
should be explored.

In the real world a $\0~\qqqq$ state may, depending on its flavor
quantum numbers, mix with $\0~\qq$ and glueball states.  It seems
natural to expect that for sufficiently heavy quarks a bound state
will remain, but only full, unquenched lattice calculations can
confirm this.

It is possible that unquenched studies of $\0~\qq$ operators may show
some corroboration of our results \cite{SCALAR,UKQCD2}.  These studies
use $\qq$ sources with dynamical fermions, but there is nothing to
stop their $\qq$ source from mixing with $\qqqq$, and allowing them to
see the $\qqqq$ bound state we have identified.  It is therefore quite
interesting that they report that the $\0$ state is significantly
lighter in unquenched calculations than in quenched ones.  However,
the calculations are still far from the continuum and chiral limits,
and it is hard to tell whether their $f_0$ will become light compared
to typical $\bar q q$ singlet states as the pion mass drops.

If light $\qqqq$ states are, in fact, a universal phenomenon, and if
the $\sigma(600)$ is predominant\-ly a $\qqqq$ object,
then the chiral transformation properties of
the $\sigma$ have to be re-examined.  The $\pi$ and
the $\sigma(600)$ are usually viewed as members of a (broken) chiral
multiplet.  In the naive $\qq$ model both $\pi$ and $\sigma$ are in
the $(\half,\half)\oplus(\half,\half)$
representation of $SU(2)_{L}\otimes SU(2)_{R}$ before symmetry
breaking.  In a $\qqqq$ model, as in the real world,
the chiral transformation properties of the $\sigma$ are not clear.

If the phenomena that we have discussed survive the introduction of
differing quark masses, then they will also have implications for
heavy quark physics. For example, there could be a $\0$ bound state
just below the decay threshold for two $D$ mesons in the charmonium spectrum.

Finally, we note that calculations similar to ours could be undertaken
in the meson-baryon sector and in other $J^{PC}$ meson channels.  It
has long been speculated that the $\Lambda(1405)$ is some sort of $KN$
bound state\cite{Pakvasa:1999zv} and $\qqqq$ states have been
postulated in other meson-meson partial waves.

\section{Acknowledgments}
  This work is supported in part by funds provided by the U.S.
Department of Energy (D.O.E.) under cooperative research agreement
\#DF-FC02-94ER40818.  The lattice QCD calculations were performed on
the SP-2 at the Cornell Theory Center, which receives funding from
Cornell University, New York State, federal agencies, and corporate
partners. The code was written by P. Lepage, T. Klassen, and M. Alford.




\IfFileExists{\jobname.bbl}{}
  {\typeout{}
   \typeout{******************************************}
   \typeout{** Please run "bibtex \jobname" to optain}
   \typeout{** the bibliography and then re-run LaTeX}
   \typeout{** twice to fix the references!}
   \typeout{******************************************}
   \typeout{}
  }

\end{document}